\magnification = \magstep 1 \baselineskip=24pt
\input amstex
\loadmsbm

\newsymbol\smallsetminus 2272
\centerline {\bf Hilbert's Sixth Problem:} 
\centerline {\bf Descriptive Statistics as New Foundations for Probability}

\centerline {Joseph F. Johnson}
\centerline {Dept. of Mathematical Sciences, Villanova Univ.}

\noindent\bf RESUMEN\rm \ 

{\narrower\smallskip\baselineskip=12pt
Hay esbozos seg\'un los cuales las probabilidades se
cuentan como la fundaci\'on de la teor\'\i a matem\'atica
de las estad\'\i sticas.  Mas la significaci\'on f\'\i sica
de las probabilidades matem\'aticas son oscuros,
muy poco entendidos.  Parec\'\i era mejor que las
probabilidades f\'\i sicas se fundaran en las
estad\'\i sticas descriptivas de datos fisicales.  Se
trata una teor\'\i a que as\'\i\  responde a una cuestiona
de Hilbert propuesta en su Problema N\'umero Seis,
la axiomatizaci\'on de la F\'\i sica. Esta est\'a basada en la
auto-correlaci\'on de los series temporales. Casi todas
de las funciones de auto-correlaci\'on de las trayector\'\i as
de un sistema din\'amico lineal (con 
un numbero bastante grande de grados de libertad) son todas
aproximadamente iguales, no importan las condiciones
iniciales, a\'un si el sistema no sea erg\'odico, como
conjetur\'o Khintchine en 1943.
\smallskip}

{\narrower\smallskip\baselineskip=12pt
Usually, the theory of probability has been made the foundation
for the theory of statistics.  But the physical significance of
the concept of probability is problematic, with no consensus. 
It would seem better to make the descriptive statistics of physical data the foundations of physical probability.  This will answer a question posed
by Hilbert in his Sixth Problem, the axiomatization of Physics.
It is based on the auto-correlation function of time series.
Almost all trajectories of a linear dynamical system (with sufficiently
many degrees of freedom) are approximately equal, no matter their initial conditions, even when the system is not ergodic, as conjectured by Khintchine in 1943.
\smallskip}

\input amstex \magnification=\magstep 1 \baselineskip=24pt
\loadmsbm

\noindent {\bf Introduction}

Hilbert's Sixth Problem [1] was the Axiomatization of Physics.
He had in mind not only the axiomatization of true physical
theories, but as well the axiomatization of false theories which
would bear an interesting resemblance to, along with 
instructive differences from, the real world. Because of the
contemporary controversies about the logical relation between Analytical Mechanics and Thermodynamics precipitated by the work of Maxwell and Boltzmann,
which involved both Poincar\'e and Zermelo, Hilbert explicitly 
pointed to the need for logical foundations for the theory of 
probability.  With Quantum Mechanics's Born's rule's having placed
probability at an even more central location in the foundations of physical theory, Hilbert's prescience is remarkable.
Readers younger than Hilbert little realize that for Hilbert and his generation, Probability was not a branch of Mathematics, it was a branch of Physics.\plainfootnote*{This point is illustrated by Corry: he found in the G\"ottingen archives the list of topics for a course Hilbert taught: ``In 1905 he taught a course on the axiomatic method where he presented for the first time a panoramic view of various physical disciplines from an axiomatic perspective: mechanics, thermodynamics, probability calculus, kinetic theory, insurance mathematics, electrodynamics, psychophysics.'' [2]}\ 
\ Hilbert realized that as a preliminary to this, one would have
to bring the theory of probability into mathematics proper by axiomatizing it in such a way as to clarify its relationships to Arithmetic or Geometry.  Fr\'echet, Wiener, and Kolmogoroff did precisely this, but Kolmogoroff well knew that this did not solve the problem of clarifying the logical foundations of what is nowadays called ``physical probability.''  He returned to this more difficult and more important part of Hilbert's Sixth Problem several times in his later career [3].

In Dirac's formulation of the axioms of Quantum Mechanics, we find the typical physicist's approach to this problem.  

{\baselineskip=14pt \narrower ``If the experiment is repeated a large number of times it will be found that each particular result will be obtained a definite fraction of the total number of times, so that one can say there is a definite probability of its being obtained any time the experiment is performed.'' [4]
\smallskip}

This is not a definition at all.  Such notions have been insightfully criticized in print by Burnside [5], Littlewood [6], and Kolmogoroff [7], all three accomplished probabilists.  In an address to a math club, Littlewood explained at length ``The Dilemma of Probability Theory.''

{\baselineskip=14pt\narrower``Now [it] cannot assert a certainty about a particular number $n$ of throws, such as `the proportion of 6's will certainly be within $p\pm\epsilon$ for large enough $n$ \dots It can only say `the proportion will lie within $p\pm\epsilon$ with at least such and such probability¿ (depending on $\epsilon$ and $n_o$)\dots 

``The vicious circle is apparent.''

``It is natural to believe that if (with the natural reservations) an act like throwing a die is repeated $n$ times the proportion of 6's will, with certainty, tend to a limit, $p$ say, as
$n\rightarrow\infty$.
(Attempts are made to sublimate the limit into some Pickwickian sense---`limit' in inverted commas.  But either you mean the ordinary limit, or else you have the problem of explaining how `limit' behaves, and you are no further.  
You do not make an illegitimate conception legitimate by putting it into inverted commas.)
\dots

``It is generally agreed that the frequency theory won't work.  But whatever 
the theory it is clear that the vicious circle is very deep-seated:  certainty 
being impossible, whatever [it] is made to state
can be stated only in terms of `probability'.  One is tempted to the extreme 
rashness of saying that the problem is insoluble (within our current 
conceptions).  More sophisticated attempts than the frequency theory have 
been made, but they fail in the same sort of way.''\smallskip}

Kolmogoroff, in a chapter [7] meant for a broad scientific audience, analyzed this logical circularity in the same way, and ten years later, having despaired of the possiblity of fixing the frequency theory, began developing his theory of algorithmic complexity as the logical foundation for probability.  

However, we can answer Littlewood's objection by, indeed, carefully defining a new kind of limit, which we will call the thermodynamic limit, which evades the logical circle of the naive frequency theory but still has physical meaning and close contact with the kind of physical content which physicists like about the frequency theory, in spite of its logical shortcomings.

The well known logician and computer scientist Prof. Jan von Plato, of Helsinki University, succeeded in giving a definition of probability for ergodic systems [8].  His definition is rather different from the one which will be given here, cannot be made to work for quantum systems [9], and because it does not use Khintchine's conjectures about the thermodynamic limit, is restricted to ergodic classical systems.

\noindent {\bf A sequence of dynamical systems}

Suppose given a sequence $M_n$ of dynamical systems, each one with $n$ degrees of freedom, and equipped with a flow $x\mapsto x_t$ and an invariant measure under the flow, $\mu_n$. 
Suppose given an observable (i.e., a measurable function) $f_n$ on each $M_n$. 
To simplify notation, if $v_n\in M_n$ is a perhaps implicitly fixed initial condition, we write $f_n(t)$ for $f_n((v_n)_t)$, the change in $f$ due to the flow.
The motivation is that we are interested in $\{M_n\}$ when in some sense they are all `the same' kind of physical system, only the number of degrees of freedom increases without bound, and $f_n$ is `the same' physical quantity, e.g., momentum.
We will, inspired by a conjecture of Khintchine's, define the limit of $M_n$ which, when it exists, is independent of the substitution of the $\mu_n$ by any other $\mu'_n$ absolutely continuous with respect to $\mu_n$.

For $f$ a measurable function of time, Wiener studied the auto-correlation function
$$\varphi_f(\tau) = \lim_{T\rightarrow\infty}{1\over2T}\int_{-T}^{T}f(t+\tau )f(t)dt.$$
When one views $f$ as an observable on $M_n$, it is a set of data, a time series, and its auto-correlation function is a descriptive statistic of this set of data.  

Wiener further defined the higher correlation functions for any positive integer $m$,
$$\varphi^m_f(\tau_1, \tau_2, \dots, \tau_m ) = \lim_{T\rightarrow\infty}{1\over2T}\int_{-T}^{T}\Pi_1^m f(t+\tau_i )dt.$$

There is no dependence on the notion of probability.
In the literature, there is a conflicting definition of the auto-correlation function $R(\tau)$ of a time series, which only applies to a time series which is not data, but really a stochastic process.  That is, suppose given a probability space $P$ with probability measure $\mu$, and for each $\alpha \in P$, suppose that $f_\alpha (t)$ is a time series in the usual sense. Then the phase auto-correlation function was defined by Khintchine several years after Wiener's work  to be $R(t,\tau) = \int_P f_\alpha(t)f_\alpha(t+\tau) d\mu ,$ and is independent of $t$ if the process is stationary (the notion of stationary seems to have been introduced by Khintchine at the same time).  The whole point here is to avoid using it, since that might seem to re-introduce the logical circle Littlewood complained about.

The whole point of thermodynamics is to convert a sequence of deterministic dynamical systems into a stochastic process by passing to whatever kind of thermodynamic limit one has defined.  Ours will be a new kind, not the same as the usual one.  Balian has called for the creation of new kinds of thermodynamic limits, each one tailored for the application at hand.

\noindent {\it Definition}.  In the setting above, the sequence $\{(M_n,\mu_n),f_n\} $ is said to {\it have a thermodynamic limit}\ if  for every choice of a compact subset $K$ of the time-axis, a positive $\epsilon$, and a positive integer $M$, there exists an integer $N$ so large that for every $n\ge N$, there exists a subset $N_n$ of $M_n$ with $\mu_n(M_n \smallsetminus 
N_n) < \epsilon$ 
such that for any two initial conditions $v$ and $w \in N_n$,
$$\vert \varphi^m_v (t_1,t_2,\dots,t_m) - \varphi^m_w (t_1,t_2,\dots,t_m) \vert < \epsilon$$ for all $t_i\in K$ and all $m < M$.  Here, $\varphi^m_v$ is the $m$-point auto-correlation function of $f_v(t) = f_n(v_t)$, and similarly for $\varphi^m_w$. 
The trajectories (or, equivalently, their initial conditions) belonging to $N_n$ are called {\it normal} and $N_n$ is called {\it a normal cell}.

It is obvious that there then exists a function $\varphi_\infty$ defined for all time
such that $\lim_{n\rightarrow\infty} \varphi_n$ converges to $\varphi_\infty$ with uniform convergence on compact sets, provided $\varphi_n$ is chosen to have an initial condition from $N_n$.  Similarly for $\varphi^m_\infty$.  The invariance under replacing $\mu_n$ by any $\nu_n$ absolutely continuous with respect to $\mu_n$ is also obvious.   

L\'evy's philosophy was that in order to study a stochastic process, it suffices to study $R(\tau)$, its auto-correlation function (in the sense of Khintchine) [10].
A Gaussian stationary centered stochastic process is determined up to equivalence by $R$.  
Wiener has also remarked [11] that even a non-Gaussian one is still determined up to some sort of equivalence by the knowledge of all its higher $m$-point auto-correlation functions $R^m$.

Since we have a set of suitable $m$-point correlation functions, we would be able to define a limit object of a sequence that has a thermodynamic limit: the stochastic process whose auto-correlation functions in the sense of Khintchine are equal to the limits of the descriptive statistics of the elements of our sequence.  But we do not need to define some sort of limit object such as this for our immediate purposes.  For now, we will regard $\phi_\infty$ and $f_\infty$ as {\it the limit}.  One would also like to define a suitable equivalence relation on the space of sequences which possess limits and study the space of equivalence classes.  Like some other Hilbert problems, the solution to the Sixth opens up many avenues for further research.

\noindent {\bf The definition of event and of probability}

The mathematical axiomization of probability theory has taught us that it is just as important to precisely specify what is an {\it event} as it is to associate a number to an event.  This, indeed, is a foundational point difficult for engineers or physicists to appreciate; they tend to feel that every subset is measurable.  In fact, the definition of Lebesgue measure formalizes an intuition about what a `physically constructible' subset of Euclidean space should be, so in a sense, non-measurable sets cannot have any physical significance.

In Quantum Mechanics, there has been the intuition that probabilities arise from the necessity of amplifying a microscopic event up to the macroscopic level (e.g., Feynman in [12]).  In Classical Mechanics, there has been the intuition that probability arises in the thermodynamic limit of deterministic systems.  (There have also been rival intuitions but we will not touch on them here.)  It follows from this that we should formally define an `event' to be something that only arises in this way, when two contrasting scales are being compared.  In particular, neither points nor subsets of a fixed $M_n$ are events.  (And for this reason, neither Lebesgue measure nor Liouville measure nor $\mu_n$ are interpreted as probability measures.) Taking our cue from Quantum Mechanics, only the result of a measurement is defined to be an event.

The quantum case was already treated, in the special case of the two slit experiment, in [13] and [14].  There, `event' was defined as the thermodynamic limit of the result of an interaction with an amplifying apparatus: in that limit, Planck's constant goes to zero and the amplifying apparatus becomes a classical system.  


In the classical case, in Statistical Mechanics, as remarked by Wiener [15], Guelfand [16], and Pauli [17], a measurement of an observable $f$ on $M$ is really a long-time average, modelled or approximated by the infinite time average 
$$\langle f \rangle _t = \lim _{T\rightarrow\infty} {1\over T}\int_0^T f(t) dt,$$ where the dependence on the initial condition $v_o \in M_n$ has been suppressed.  The same applies to any function of $f$, for example, the variance $f^2$.\plainfootnote*{ In the author's view, and in the views just cited, time averages model measurements and phase averages model probabilities.  Prof. von Plato, following Einstein and in agreement with Landau, defines probabilities as infinite time averages.}\  However, this dependence on the initial condition prevents us from turning this idea into an exact definition of {\it event}, and this is the reason we pass to the thermodynamic limit.  As the number of degrees of freedom grows without bound, almost all initial conditions give approximately the same answer, the expectation of $f$, and $\varphi_\infty (0)$ for the variance.  We also obtain all the higher moments of the limit of $f$, and so a random variable $f_\infty$ can be rigorously defined (as usual, its probability space $M_\infty$ is taken to be the unit interval [0,1] with Lebesgue measure).   Physically, $f_\infty$ is an idealisation with properties which are good approximations to the vast majority of the $\langle f_n \rangle_t$, $\langle f_n^2 \rangle_t$, etc., each of which is a descriptive statistic of some concrete data.

Suppose given a sequence $\{(M_n,\mu_n),f_n\}$ which has a thermodynamic limit, with its associated $\varphi_{\infty}$, $f_\infty$, etc., as above.  


%
%
\noindent {\it Definition}.  Let the probability space $P$ be the direct image under $f_\infty$ of the probability space $(M_\infty, dx)$.  Then the {\it events} of the thermodynamic limit of $\{(M_n,\mu_n),f_n\}$ are the measurable subsets of $P$ and the {\it probability} of an event $F$ is its measure.


The definition of limit we have introduced is modelled closely on the equilibrium statistical mechanics and work of Ford, Kac, and Mazur [18].  For this reason, the measurement yields one value with probability unity, because the system is in equilibrium.  

In fact, this limit was tailor-made for measurements of $f$, but it will apply as well to any function of $f$.  If $f$ models the coin-toss (or cast of a die) dilemma of Littlewood, \plainfootnote*{If one were to construct the obvious stochastic process from the idea of repeated coint-tossing, the process would not be stationary in continuous time.  But in our use of descriptive statistics, there is no assumption of stationarity.}\ then $f$ will be assumed to take only the values $\pm\frac12$, and be centered.  Composing $f$ with the indicator function of a small neighbourhood of $\frac12$, we get $g$ (or, just put $g=f+\frac12$). Then $\langle g \rangle_t= $\ the frequency of heads.\plainfootnote\dag{For us, frequency is not equal to probability.  What is measured is frequency.  The frequency is related in a subtle way to the probability, just as time averages are related to phase averages.}\   
This frequency might be zero, if the initial condition is perverse.
Putting $g_\infty = f_\infty+\frac12$, all the moments of $g_\infty$ follow from those of $f_\infty$.  In particular, the expectation of $g_\infty$, which is our definition of the probability that $f$ takes the value ``heads,'' depends only on the equivalence class of the sequence $\{(M_n,\mu_n),f_n\}$.

The physical meaning is that if the sequence was defined shrewdly, then it is a good approximation to $\langle g_{6.2\cdot10^{23}}\rangle_t$ (a physically meaningful function) unless the initial condition does not belong to $N_n$, which is a determinate statement with concrete physical meaning.
Of course the limit of the sequence does not change, and hence 
$\langle g_\infty \rangle_\mu$ does not change, if any finite number of $M_n$ are replaced by ridiculous counterfeits, and this includes 
$M_{6.2\cdot10^{23}}$.  In this case, the statement will be useless for any practical purpose, but still physically meaningful.  The same applies if the initial condition is, in fact, outside of $N_{6.2\cdot10^{23}}$.  The statement will be meaningful but useless for this particular case.  Many have already suspected that the true meaning of probability is an approximate one with a certain range of validity, and when used outside the limits of that range, will lead to paradoxes or practically useless statements.
And the point of the Hilbert problem is only to tidy the logical structure
of probability statements, not to impose a tidiness on the world that does not exist. 

Ever since the work of Wiener, physicists and engineers have had the intuition that a time series whose auto-correlation function has an absolutely continuous power spectrum is ``random.''  This can be made precise in the context of our definition.  If the coin-tosses result from $\{(M_n,\mu_n),f_n\}$, as above, then we can use the auto-correlation of a sequence of unit pulses as a measure of how random the sequence is.  If its auto-correlation function is {\it normal}, i.e., approximately equal to that of all the others from $N_n$, then the sequence is approximately random.  Thus, the auto-correlation function can be used instead of ideas of algorithmic complexity.

The assertion that the probabilities in the thermodynamic limit are good approximations to the real situation of $M_{6.2\cdot10^{23}}$ is testable, by experiment.  In principle, one should, in many concrete cases of this limit, be able to calculate how large $n$ has to be. If the predictions based on calculations using the limit are falsified by an experimental run, then $v_o\notin N_n$.  That said, the practical purpose of using thermodynamic limits is precisely to avoid having to make calculations about $M_n$, which are practically impossible, substituting for them calculations about $M_\infty$, which are easier.  

\noindent {\bf A class of examples}

We will show that this definition is not vacuous by studying an interesting class of examples: Hamiltonian systems of linearly coupled harmonic oscillators.  These systems are completely integrable, but in the limit, they exhibit the kind of very very weak ergodicity conjectured by Khintchine in 1943
[19] for a (hopefully)  much larger class of dynamical systems (he did not concretely specify which class).  The first point is that the systems are simple enough that the calculations for $M_n$ can be carried out.
The second point is that ergodicity is usually associated with non-linearity, but here are linear systems which on the macro-level are practically indistinguishable from ergodic systems.

The third point is that from the standpoint of the foundations of Physics, only Quantum Mechanics is truly important, not Classical Mechanics, and quantum systems are linear Hamiltonian systems.  So we will study the general class of linearly coupled harmonic oscillators as in [20].

Obviously not every sequence of systems $M_n$, even if possessing a limit, will exhibit weakly ergodic behaviour even if $n$, the dimension of the space, increases without bound.  The intuition from equilibrium statistical mechanics is that each $M_n$ must be composed of many identical parts (or, more generally, a fixed number of different types of parts with the number of parts of the same type increasing without bound), and there must be a coupling between the parts.  Furthermore, a natural hypothesis to make is that the interaction between part $i$ and part $j$ only depends on the relative situation of $i$ and $j$, so that if $k$ and $l$ constitute a parallel pair, their interaction term should be the same.  This leads naturally to the study of an interaction matrix $A_n$ which is cyclic (and, of course, symmetric). 

We will generalize the result of [20], which in turn was a generalization of the results of Ford, Kac, and Mazur [18].  The main point here is only to show how the new definition of probability and event applies in this situation.  The main interest is that the same kind of definition of {\it probability} and {\it event} works for classical physics as was used earlier, in [13] and [14], for the quantum mechanical measurement of a two-state system by an amplifying apparatus in a state of negative temperature.  The second point of interest is that we will have introduced the notion of probability without relying on imposing a particular probability distribution on $M_n$.  This opens the way, in the future, to studying systems in a negative temperature state, where the usual notion of probability distribution cannot be used.

\noindent {\it Notation.} \ If $n$ is even, choose $M_n$ to be the same as $M_{n-1}$.
From now one, assume $n$ is odd, and equal to $2N+1$.  All 
indices will run from $-N $ to $N$, except angles, which will run from epsilon above $-\pi$ to epsilon below $\pi$: we put $\theta_l = { 2\pi l \over n }$ for $l = -N, \dots , N $.

 $M_n$ is a Hamiltonian dynamical system (or, rather, the restriction of one to a surface of constant energy, see later) with canonical co-ordinates $p_i$, $q_i$ and   Hamiltonian 
$H_n$
$$H_n=  \sum_{i=-N}^N {p_i^2\over 2m} + 
{1\over2}  (q_{-N}, q_{-N+1}, \dots  q_N) A 
 \pmatrix  q_{-N} \\
 q_{-N+1}\\ \vdots  \\ q_N \endpmatrix$$
where A is a symmetric $n\times n$ square real matrix with positive eigenvalues $\omega^2_l$
satisfying
$$(A)_{ml}={1\over n+1}\sum_{k=-N}^{N}\omega_k^2e^{{2\pi {\sqrt -1} \over n+1} k(m-l)}.$$
\noindent This is obviously symmetric if we make a simple assumption on the $\omega_l$'s.

We have 
$$p_o(t)={1\over n}
\left\{\sum_k \sum_l \cos (\omega_lt )\zeta^{-lk}p_k(0) 
-\sum_k \sum_l \omega_l\sin (\omega_lt )\zeta^{-lk}q_k(0) 
\right\}.$$
Putting $\widehat p (k) = \sum_i \zeta^{-ik} p_i(0)$ and similarly for $\widehat q$, this 
becomes
$$ p_o(t) = {1\over n} \left\lbrace
\sum_k \widehat p(k) \cos (\omega_k t)
- \sum_k \widehat q(k) \omega_k \sin (\omega_k t)\right\rbrace . \tag 1 $$

Furthermore, the auto-correlation function of $p_o$ is
$$\varphi (\tau) = \sum_k \frac12 \left({1 \over 2N+1}\right)^2
(\vert \widehat p(k) \vert^2 +
 \vert \omega_k \widehat q(k) \vert^2)
\cos ( \omega_k t) ,\tag 2 $$ 
and the higher auto-correlation functions vanish for an odd number of points and, for an even number of points, are trigonometric polynomials with more or less the same coefficients.



We will take $p_o$ as our observable $f_n$,
and the restriction of Liouville measure to any surface of constant energy $E$ as our invariant measure $\mu_n$.  The dynamical system $M_n$ will be the surface of constant energy.
The energy level $E_n$ is defined for traditional reasons, and to make the 
comparison with traditional results convenient, to be that energy level which 
is most probable according to the Maxwell distribution: it is $n\over kT$, where $k$ is Boltzmann's constant and $T$ is the absolute temperature in degrees Kelvin.

To implement the notion that the $M_n$ are the same but different, we will suppose that their eigenvalues are taken from the same function $\omega$ but evaluated at different points.  Suppose that we know the eigenvalues $\omega_l$ for the real system $M_{6.2\cdot 10^{23}} $ which we are given.  Regarding $\omega_l$ as a function of $\theta_l$, write it as $\omega(\theta_l) = \omega_l$.  But now regard $\omega$ as a continuous function on $(-\pi, \pi) $ by interpolating the given values in some sensible fashion.  (One that makes intuitive physical sense.)

For any $n$, define the Hamiltonian of $M_n$ by putting $\omega_s=\omega({ 2 \pi s \over n})$.  
Then the sums in Equation 1 become Riemann sums for the improper integrals
$$ {1\over 2\pi} 
\int_{-\pi}^\pi \widehat p(\theta) \cos (\omega(\theta) t) d\theta 
- \int_{-\pi}^\pi \widehat q(\theta) \omega(\theta) \sin (\omega(\theta) t) d\theta .$$

Now the same methods of proof of the theorem of [20] show that 

\noindent {\it Theorem}.  Suppose that $\omega$ is a continuous function on $(-\pi, \pi)$ such that the Riemann integrals 
$$\int_{-\pi+\delta}^{\pi-\delta}  \omega(\theta)\cos (m\theta) d\theta$$ 
converge for every $m$ and every small positive $\delta$.
Using $\omega$, define $\{(M_n,\mu_n), f_n \}$ as above.  Then this sequence has a thermodynamic limit, and 
$$\varphi_\infty(\tau) = {1 \over 2\pi} \int_{-\pi}^\pi
\cos( \omega(\theta) \tau ) d\theta.$$ 

\noindent {\it Corollary}.  In fact, since the coefficients are more or less the same for all the higher multi-point auto-correlation functions as they are for the ordinary one $\varphi$, the proof shows more.  It shows the uniformity in $M$ of our estimates, and hence, this sequence satisfies a stronger condition than is necessary for the definition of limit: the conclusion holds for all $m$ simultaneously.

We omit the details of the proof of the corollary.

\centerline{\bf References}

\noindent [1] D. Hilbert, ``Probl\`emes Math\'ematiques,'' {\it 
Compte Rendu du Deuxi\`eme Congr\`es International des Mathematiciens Tenu \`a Paris du 6 au 12 ao\^ut 1900}, Paris, 1902, pp.~81--83.

\noindent [2] L. Corry,
``On the Origins of Hilbert's Sixth Problem: Physics and the
Empiricist Approach to Axiomatization.'' In:
{\it Proceedings of the International Congress
of Mathematicians, Madrid, Spain, 2006}, vol. 3, pp. 1697--1718.

\noindent [3] A. Kolmogoroff, talk given at the 1970 ICM at Nice, France, unpublished. 

\noindent [4] P. A. M. Dirac, {\it Principles of Quantum Mechanics}, 1st ed., Oxford, 1930, p.~10 

\noindent [5] W. Burnside, ``On the Idea of Frequency,'' {\it Proc. Camb. Phil. Soc.} {\bf 22} (1925), 726.

\noindent [6] J. Littlewood, ``The Dilemma of Probability Theory,'' p.~32. In: J. Littlewood, {\it A Mathematician's Miscellany}, Cambridge, 1956.

\noindent [7] A. Kolmogoroff, ``Teoriya Veroyatnostey,'' pp.~261--262.  In: A. Alexandroff, A. Kolmogoroff, and M. Lavrentieff, eds., {\it Matematika, yeyo Soderzhanie, Metody, i Znacheniya}, 2nd ed., vol. 2, Moscow, 1956, pp.~252--284.

\noindent [8] J. von Plato, ``Ergodic Theory and the Foundations of Probability.'' In: B. Skyrms and W. Harper, eds., {\it Causation, Chance, and Credence, Proceedings of the Irvine Conference on Probability and Causation}, vol.\ 1, Dordrecht, 1988, pp.\ 257--277.

\noindent [9] J. von Plato, personal communication.

\noindent [10] P. L\'evy, \it Processes Stochastiques et Mouvement Brownien, \rm Paris, 1948. 

\noindent [11] N. Wiener, {\it Extrapolation, Interpolation, and Smoothing of Stationary Time Series: With Engineering Applications}, Cambridge, Mass., 1949, p.~18.

\noindent [12] 
R. Feynman and A. Hibbs, {\it Quantum Mechanics and Path Integrals}, New York, 1965, p.~22.

\noindent [13] J. Johnson, ``Statistical Mechanics of Amplifying Apparatus.'' In: 
S.\ Catto and B. Nicolescu, eds., \it Proc.\ VIII International Wigner Symposium, New York City, 2003. \rm

\noindent [14] J. Johnson, ``Thermodynamic Limits, Non-Commutative Probability, and Quantum Entanglement.'' In: P. Argyres \it et al\rm., eds., {\it Quantum Theory and Symmetries, Proceedings of the Third International Symposium, Cincinnati, 2003}, Singapore, 2004, pp.~133--143.

\noindent [15] P. Masani and N. Wiener, ``Non-linear
Prediction.'' In: U.\ Grenander, ed., {\it Probability and Statistics, The Harald Cramer Volume},
Stockholm, 1959, p.~197.

\noindent [16] I. Guelfand and N. Vilenkin, {\it Les distributions tome 4: applications de l'analyse harmonique}, Paris, 1967, pp.~237f.

\noindent [17] W. Pauli, {\it Pauli Lectures on Physics, volume 4, Statistical
Mechanics}, Cambridge, Mass., 1973, pp.~28f.

\noindent [18] G. Ford, M. Kac, and P. Mazur, ``Statistical Mechanics of Assemblies of Coupled Oscillators,'' \it J.\ Math.\ Phys.\ {\bf6} \rm(1965), pp.~504--515.

\noindent [19] A. Khintchine, {\it Matematichiskie Osnovaniya Statisticheskoi Mekhaniki}, Moscow, 1943. English translation, {\it Mathematical Foundations of Statistical Mechanics}, New York, 1949. 

\noindent [20] J. Johnson, ``Some Special Cases of Khintchine's Conjectures in Statistical Mechanics: Approximate Ergodicity of the Auto-Correlation Functions of an Assembly of Linearly Coupled Oscillators,'' \it Revista Investigaci\'on Operacional \rm \ {\bf33}, no.~3, pp.~99--113.

\end



\Wiener\foo 3 [Masani, Wiener, ``Non-linear
Prediction,'' in [\it Probability and Statistics, The Harald Cramer Volume],
ed.\ U.\ Grenander, Stockholm, 1959, p.\ 197: ``As indicated by von Neumann \dots
in measuring a macroscopic quantity $x$ associated with a physical or biological
mechanism\dots each reading of $x$ is actually the average over a time-interval
$T$ [which] may appear short from a macroscopoic viewpoint, but it is large
microscopically speaking.  That the limit $\overline x$, as $T \rightarrow
\infty$, of such an average exists, and in ergodic cases is independent of the
microscopic state, is the content of the continuous-parameter $L_2$-Ergodic
Theorem.  The error involved in practice in not taking the limit is naturally to
be construed as a [\it statistical dispersion] centered about $\overline x$.''

Pauli, [\it Pauli Lectures on Physics, volume 4, Statistical
Mechanics], Cambridge, Mass., 1973, p.\ 28f., ``What is observed macroscopically
are time averages\dots ''] (who was a logician), interpret the time average of a
functional as `measurement.'  

`Disons quelques mots des origines physiques de la notion de distribution aleatoire.
La notion habituelle de fonction aleatoire repose sur l'hypthese qu'a chaque instant $t$,
il est possible de mesurer les valeurs de la fonction aleatoire
independamment de ses valeurs a d'autres instants.  Mais, en fait,
toute mesure reelle s'effectue a l'aide d'un appareil ayant une certaine inertie.
Donc, les indications de l'appareil ne fournissent pas les valeurs de la grandeur 
aleatoire $\xi(t)$ a l'instant $t$, mais une certaine grandeur moyenne : 
$\Phi(\varphi) = \int \varphi(t) \xi(t) dt$,
ou $\varphi(t)$ est une functions caracteristique de l'appareil.' 
I. Guelfand et N. Vilenkin, Les distributions tome 4 applications de l'analyse harmonique Paris 1967, p.\ 237f

Ever since the work of Wiener, physicists and engineers have had the intuition that a time series whose auto-correlation function has an absolutely continuous power spectrum is ``random.''  This can be made precise in the context of our definition.  If the coin-tosses result from $\{(M_n,\mu_n),f_n\}$, as above, then we can use the auto-correlation of a sequence of unit pulses as a measure of how random the sequence is.  If its auto-correlation function is {\it normal}, i.e., approximately equal to that of all the others from $N_n$, then the sequence is approximately random.  Thus, the auto-correlation function can be used instead of ideas of algorithmic complexity.
\noindent [20] O. Lanford III, Entropy and Equilibrium States in Classical Statistical Mechanics, in A. Lenard, ed., \it Statistical Mechanics and Mathematical Problems, Battelle Seattle 1971 Rencontres\rm, Lecture Notes in Physics vol.\ 20, Berlin, 1973. 

Littlewood at length.
`Mathematics \dots has no grip on the real world;
if probability is to deal with the real world it must contain elements 
outside mathematics, the \it meaning \rm of `probability' must relate 
to the real world; and there must be one or more `primitive' propositions
about the real world, from which we can then proceed deductively (\it i.e.
\rm mathematically).  
We will suppose 
(as we may by lumping several primitive propositions together) 
that there is just one primitive proposition, 
the `probability axiom', and 
we will call it `\it A\rm' for short.\smallskip}

{\baselineskip=14pt\narrower`\dots the `real' probability problem; 
what are the axiom \it A \rm and the meaning of `probability' to 
be, and how can we justify \it A\rm? 
It will be instructive to consider the attempt called the `frequency theory'.  It is 
natural to believe that if (with the natural reservations) an act like throwing 
a die is repeated $n$ times the proportion of 6's will, \it with certainty\rm, 
tend to a limit, $p$ say, as $n \rightarrow \infty$.  (Attempts are made to 
sublimate the limit into some Pickwickian sense---`limit' in inverted commas.  
But either you \it mean \rm the ordinary limit, or else you have the problem 
of explaing how `limit' behaves, and you are no further.  You do not make an 
illegitimate conception legitimate by putting it into inverted commas.)  If we 
take this proposition as `\it A\rm' we can at least settle off-hand the other 
problem, of the \it meaning \rm of probability,  we can define its measure for 
the event in question to be the number $p$.  But for the rest this \it A \rm 
takes us nowhere.  
\dots Now an \it A \rm cannot assert a \it certainty \rm about a 
particular number $n$ of throws, such as `the proportion of 6's will \it 
certainly \rm be within $p\pm\epsilon$ for large enough $n$ (the largeness 
depending on $\epsilon$)'.  
It can only say 'the proportion will lie between $p\pm\epsilon$  \it with at 
least such and such probability (depending on $\epsilon$ and $n_o$) whenever 
$n>n_o$'\rm.  The vicious circle is apparent.  We have not merely failed to 
\it justify \rm a workable 
\it A\rm; we have failed even to \it state \rm one which would work if its 
truth were granted.
It is generally agreed that the frequency theory won't work.  But whatever 
the theory it is clear that the vicious circle is very deep-seated:  certainty 
being impossible, whatever \it A \rm is made to state
can be stated only in terms of `probability'.  One is tempted to the extreme 
rashness of saying that the problem is insoluble (within our current 
conceptions).  More sophisticated attempts than the frequency theory have 
been made, but they fail in the same sort of way.\smallskip}  

     Feynman, famously,$^{8}$ did not think that Quantum Measurement was an 
important problem, but he did think that perhaps a little more could be said 
about it (p.\ 22): 

{\narrower\smallskip\baselineskip=12pt``We and our measuring instruments are part of nature and so are, in principle, described by an amplitude function satisfying a deterministic equation.  
Why can we only predict the probability that a given experiment will lead to a definite result?
From what does the uncertainty arise?  Almost without a doubt it arises from the need to amplify the effects of single atomic events to such a level that they may be readily observed by large systems. \smallskip}  

{\narrower\hskip-10pt\baselineskip=12pt ``\dots In what way is only the probability of a future event accessible to us, whereas the certainty of a past event can often apparently be asserted?\dots Obviously, we are again involved in the consequences of the large size of ouselves and of our measuring equipment.  
The usual separation of observer and observed which is now needed in analyzing measurements in quantum mechanics should not really be necessary, or at least should be even more thoroughly analyzed.  
What seems to be needed is the statistical mechanics of amplifying apparatus. \smallskip}